\documentclass[twocolumn]{article}

\usepackage[T1]{fontenc}

\usepackage{geometry}
\usepackage{setspace}

\usepackage{hyperref}

\usepackage[style = chem-acs, articletitle=true]{biblatex}
\DeclareSourcemap{
  \maps[datatype=bibtex]{
    \map{
      \step[fieldsource=journal, match=\regexp{Physical\s+Review\s+Letters}, replace={Phys. Rev. Lett.}]
      \step[fieldsource=journal, match=\regexp{Journal\s+of\s+American\s+Chemical\s+Society}, replace={J. Am. Chem. Soc.}]
      \step[fieldsource=journal, match=\regexp{Nature\s+Communications}, replace={Nat. Commun.}]
      \step[fieldsource=journal, match=\regexp{Chemical\s+science}, replace={Chem. Sci.}]
      \step[fieldsource=journal, match=\regexp{Journal\s+of\s+the\s+American\s+Chemical\s+Society}, replace={J. Am. Chem. Soc.}]
      \step[fieldsource=journal, match=\regexp{Nature\s+Chemistry}, replace={Nat. Chem.}]
      \step[fieldsource=journal, match=\regexp{ACS\s+Catalysis}, replace={ACS Catal.}]
      \step[fieldsource=journal, match=\regexp{Chemical\s+Society\s+Reviews}, replace={Chem. Soc. Rev.}]
      \step[fieldsource=journal, match=\regexp{Nature\s+Catalysis}, replace={Nat. Catal.}]
      \step[fieldsource=journal, match=\regexp{Angewandte\s+Chemie\s+International\s+Edition}, replace={Angew. Chem. Int. Ed.}]
      \step[fieldsource=journal, match=\regexp{The\s+Journal\s+of\s+Chemical\s+Physics}, replace={J. Chem. Phys.}]
      \step[fieldsource=journal, match=\regexp{Angewandte\s+Chemie}, replace={Angew. Chem.}]
      \step[fieldsource=journal, match=\regexp{nature\s+communications}, replace={Nat. Commun.}]
      \step[fieldsource=journal, match=\regexp{(The\s+)?Journal\s+of\s+Physical\s+Chemistry\s+Letters}, replace={J. Phys. Chem. Lett.}]
      \step[fieldsource=journal, match=\regexp{Journal\s+of\s+Chemical\s+Physics}, replace={J. Chem. Phys.}]
      \step[fieldsource=journal, match=\regexp{Physical\s+Review\s+Research}, replace={Phys. Rev. Res.}]
      \step[fieldsource=journal, match=\regexp{Chemical\s+Reviews}, replace={Chem. Rev.}]
      \step[fieldsource=journal, match=\regexp{Journal\s+of\s+Physics\s+B:?\s*Atomic,\s*Molecular\s+and\s+Optical\s+Physics}, replace={J. Phys. B: At. Mol. Opt. Phys.}]
      \step[fieldsource=journal, match=\regexp{(Proceedings\s+of\s+the\s+National\s+Academy\s+of\s+Sciences(\s+of\s+the\s+United\s+States\s+of\s+America)?|PNAS)}, replace={Proc. Natl. Acad. Sci. U.S.A.}]
    }
  }
}

\usepackage{graphicx}
\usepackage{float}
\usepackage{caption}
\newfloat{scheme}{htbp}{los}
\floatname{scheme}{Scheme}
\floatname{chart}{Chart}
\newfloat{graph}{htbp}{loh}
\usepackage{abstract}

\usepackage{chemformula} 
\usepackage[version = 4]{mhchem}

\usepackage{authblk}
\author[1,2]{Jillian G. Hoffstadt}
\author[1,2]{Davis M. Welakuh}
\author[1,2,3]{Norah M. Hoffmann}
\affil[1]{Department of Chemistry, New York University, New York, NY 10003, USA}
\affil[2]{Simons Center for Computational Physical Chemistry, New York University, New York, NY 10003, USA}
\affil[3]{Department of Physics, New York University, New York, NY 10003, USA}

\title{Characterizing Reaction Types through Linear Free-Energy Relationships in Cavities} 
\date{*Email: norah.hoffmann@nyu.edu}

\begin{document}

\twocolumn[
  \begin{@twocolumnfalse}
    \maketitle

    \begin{onecolabstract}
      Polaritonic chemistry holds the promise of controlling chemical reactivity through confined light--matter interactions, yet how reactions respond inside optical cavities remains debated. Here, we take a more zoomed-out view, inspired by electrostatic catalysis. We extend nonempirical linear free-energy relationships to cavity-modified reactions and show that their behavior falls into four distinct reaction types set by simple response-crossing rules. Reactions can follow conventional Bell--Evans--Polanyi (BEP) or counterintuitive Opposite-BEP behavior regardless of cavity tuning, or switch between the two as the cavity frequency is varied. We establish this picture analytically and demonstrate it in the Shin--Metiu model and in four molecular reactions treated with \textit{ab initio} electronic-structure methods. Even at this deliberately coarse-grained level, cavity-modified reactivity already shows qualitatively different and strongly counterintuitive behavior, including reactions that switch character with cavity frequency. Yet this complexity can still be reduced to a simple competition between permanent-dipole and polarizability responses. This reaction-type classification therefore provides a new route to organize and anticipate cavity-modified reactivity.
    \end{onecolabstract}

    \vspace{2em}
  \end{@twocolumnfalse}
]

\section*{Introduction}

A central challenge in chemistry is controlling individual molecules to manipulate material properties and chemical reactions. In recent years, various strategies have emerged to exert such control. One elegant example is electrostatic catalysis, where external electric fields are used to catalyze or inhibit reactions, providing a means to manipulate reaction kinetics without altering chemical composition \autocite{shaik2016, che2018, ciampi2018, foroutan2014, lau2017, gorin2012, gorin2013, welborn2018, shaik2020, joy2020}. A related, more nascent field is polaritonic chemistry, which uses the confined, structured electric fields of optical cavities to modify chemical behavior \autocite{thomas2016, thomas2019, sau2021, hirai2020prins, ahn2023science, lather2019, vergauwe2019, pang2020,lather2022, fidler2023, chen2024, yin2025}. When molecules are strongly coupled to a cavity, their vibrational or electronic excitations hybridize with the photon modes, forming new eigenstates, the upper and lower polaritons, that can potentially open up new reaction pathways. This approach holds the promise of controllable cavity-induced catalysis, where reactivity is tuned not through heat or solvents, but by the cavity field arising from the cavity confinement. However, despite its promise, and in contrast to electrostatic catalysis, polaritonic chemistry remains marked by contradictions and open questions \autocite{hoblos2025, lamb}.

To date, only a small number of chemical reactions have been studied inside optical cavities, either in dark-cavity settings \autocite{thomas2016, thomas2019, lather2019, hirai2020prins, sau2021} or under external driving \autocite{pang2020, chen2022, yang2023, xiang2020}, and among those, the results remain inconsistent. Some report increased reaction rates \autocite{lather2019, lather2022}, others report rate suppression \autocite{thomas2016, thomas2019, vergauwe2019, hirai2020prins}, and some observe no change at all \autocite{fidler2023, chen2024}. In response, many theoretical explanations have emerged, ranging from reshaping potential energy surfaces \autocite{pavosevic2023nature, climent2019plasmonic, schaefer2021, galego2019, mandal2023} and modifying reaction kinetics \autocite{lindoy2023dynamics, lindoy2022dynamics, limandal2021} to collective energy-transfer effects \autocite{sharma2024}, analytical models \autocite{lai2026}, and more recent ideas involving local polarization \autocite{horak2025}, hot spots\autocite{wickramasinghe2026mode}, and polarization-glass physics \autocite{sidler2026, hasyim2026}. While these directions have led to major advances and established an important research area at the intersection of materials science and cavity quantum electrodynamics, the microscopic origin of polaritonic chemistry -- both in dark and driven cavities -- remains unclear.

\begin{figure}[h!]
\centering
\includegraphics[width=\columnwidth]{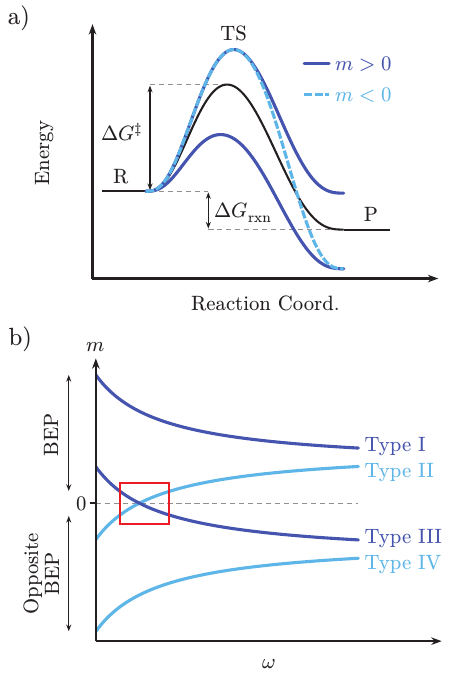}
\caption{\label{fig:figure0}
\textbf{(a)} Schematic of a reaction pathway with reactant (R), transition state (TS), and product (P) (black line), with the activation energy $\Delta G^{\ddagger}=E_{\mathrm{TS}}-E_{\mathrm{R}}$ and reaction energy $\Delta G_{\mathrm{rxn}}=E_{\mathrm{P}}-E_{\mathrm{R}}$. Field-modified reaction pathways are shown for Bell--Evans--Polanyi (BEP) behavior with $m>0$ (dark blue) and Opposite-BEP behavior with $m<0$ (light blue). \textbf{(b)} Schematic of the four reaction types in a cavity: Type I (dark blue) remains at $m>0$, Type IV (light blue) remains at $m<0$, while Type II (light blue) and Type III (dark blue) cross $m=0$ upon tuning the cavity frequency $\omega$, switching between BEP and Opposite-BEP behavior.
}
\end{figure}

In this work, drawing on established concepts from electrostatic catalysis, we take a more zoomed-out view of polaritonic chemistry and, rather than asking about its microscopic origin, ask what types of reaction behavior can generally emerge. We show that cavity-modified reactions do not respond in a universal way, but instead fall into four distinct reaction types, ranging from conventional to counterintuitive kinetic–thermodynamic behavior. These reaction types can be predicted from molecular response properties, and a single reaction can even switch between them as the cavity frequency is tuned. To this end, we extend the recently introduced nonempirical linear free energy relationships (LFERs) in static electric fields to cavity modified systems \autocite{hoffmann2022linear}. LFERs linearly connect the kinetics of a reaction to its thermodynamics
\begin{equation}
\Delta\Delta G^\ddag = m \Delta\Delta G_{\mathrm{rxn}} ,
\label{lin}
\end{equation}
where $\Delta\Delta G^\ddag$ is the change in activation energy and $\Delta\Delta G_{\mathrm{rxn}}$ is the change in reaction energy (Fig.~\ref{fig:figure0}(a)). This relationship, if satisfied, is a powerful tool to predict reactions, as it bypasses the need for detailed knowledge of the transition state, information that is often difficult or even impossible to obtain. While many families of reactions are empirically (i.e., without an atomistic expression for $m$) known to follow such relationships, previous work demonstrated that in the presence of a static electric field, the proportionality constant $m$ can be derived from molecular properties: ~\autocite{hoffmann2022linear}
\begin{equation}
m = \frac{\Delta\mu^\ddag}{\Delta\mu_{\mathrm{rxn}}},
\label{prefactor}
\end{equation}
with $\Delta\mu^\ddag = \mu_{\mathrm{TS}} - \mu_{\mathrm{R}}$ and $\Delta\mu_{\mathrm{rxn}} = \mu_{\mathrm{P}} - \mu_{\mathrm{R}}$, where $\mu$ denotes the field-free dipole moments of the transition state (TS), reactant (R), and product (P). This yields a nonempirical LFER, where knowledge of the field-free dipoles enables the prediction of reaction behavior under applied electric fields. 

\begin{figure*}[h!]
    \centering
    \includegraphics[width=\linewidth]{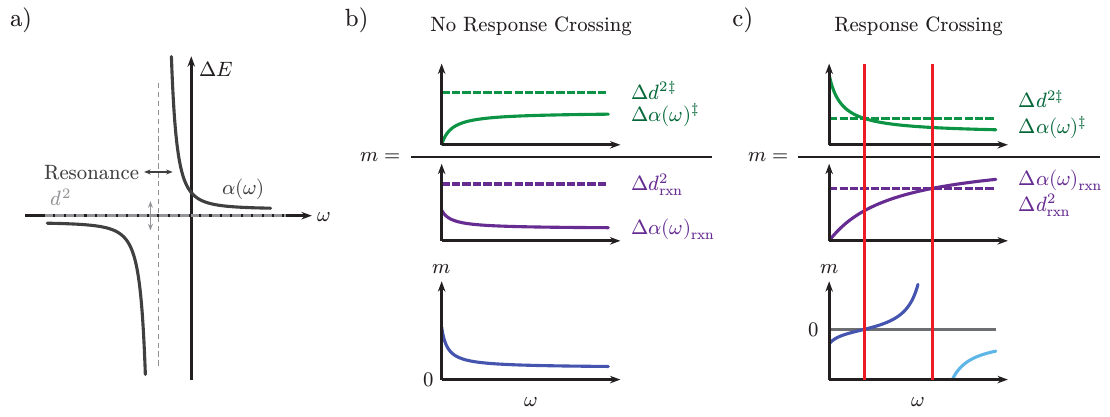}
    \caption{\textbf{Analytical structure of the cavity-modified LFER prefactor $m$.} \textbf{(a)} Schematic decomposition of the cavity-induced energy shift into a frequency-independent permanent-dipole contribution, $d^2$ (light gray), which sets the vertical offset, and a frequency-dependent polarizability contribution, $\alpha(\omega)$ (dark gray), which gives rise to the characteristic pole-like $1/x$ structure. \textbf{(b)} Schematic numerator (top, green) and denominator (middle, purple) contributions entering $m$ in the absence of a response crossing. The dashed lines indicate the permanent-dipole contributions, while the solid lines show the corresponding frequency-dependent polarizability contributions. Since neither contribution crosses its corresponding permanent-dipole term over the cavity-frequency range considered, the signs of the numerator and denominator remain unchanged and $m$ does not cross zero (bottom). \textbf{(c)} Schematic illustrating response crossings in the numerator and denominator of $m$. A response crossing in the numerator (first red line) changes the sign of the numerator and produces a zero crossing in $m$ (bottom), whereas a response crossing in the denominator (second red line) causes the denominator to vanish and produces a pole in $m$.}
    \label{fig:figure1}
\end{figure*}

Typically, the prefactor $m$ is positive, indicating that more thermodynamically favorable reactions proceed faster, consistent with the Bell-Evans-Polanyi (BEP) principle (Fig.~\ref{fig:figure0}(a), dark blue). However, in electric fields, the dipole-based expression for $m$ allows for negative values, giving rise to a counterintuitive scenario in which a reaction becomes slower while becoming more thermodynamically favorable, or vice versa (Fig.~\ref{fig:figure0}(a), light blue). This counter-intuitive case was first reported in Ref. \autocite{hoffmann2022linear} and we will refer to this scenario as the \textit{Opposite BEP}.

\section*{Results and Discussion}

Here, we derive the cavity analogue of the LFER from the Pauli-Fierz Hamiltonian \autocite{craig1998, rokaj2017, schaefer2018, flick2017} using an adjusted second-order perturbation theory for the light–matter interaction \autocite{climent2019plasmonic, schiff1968quantum}. This adjustment retains the permanent-dipole contribution, which otherwise cancels at second order, allowing us to capture its interplay with the frequency-dependent polarizability. This is central to our zoomed-out approach of connecting cavity-modified reactivity directly to accessible molecular response properties. Details of the Hamiltonian, the perturbative derivation, and benchmarks of the adjusted expression against full perturbation theory and exact solutions are provided in Supporting Information Sec.~1. The resulting adjusted energy shift of the molecular state $\psi_n$ is given by

\begin{align}
\Delta E_{n} = \frac{1}{2} &\sum_{m}^{\infty} |\langle \psi_n | \boldsymbol{\lambda} \cdot \boldsymbol{\hat{d}} | \psi_m \rangle |^2 \nonumber \\
&- \sum_{m\neq 0 }^{\infty} \frac{\hbar \omega}{2} \frac{|\langle \psi_n | \boldsymbol{\lambda} \cdot \boldsymbol{\hat{d}} | \psi_m \rangle |^2}{\varepsilon_m - \varepsilon_n + \hbar \omega} ,
\label{eq1}
\end{align}
where $\boldsymbol{\hat{d}}$ is the dipole operator, $\psi_n$ and $\psi_m$ are the unperturbed molecular eigenstates, $\varepsilon_n$ and $\varepsilon_m$ their corresponding energies, $\omega$ is the cavity mode frequency, and $\boldsymbol{\lambda}$ is the light-matter coupling vector. Eq. (\ref{eq1}) can be recast in a more compact form as
\begin{equation}
\Delta E_n = \frac{1}{2} \boldsymbol{\lambda}^2 \cdot \left( \boldsymbol{\mu}_n^2 - \omega\boldsymbol{\alpha}(\omega)_n \right),
\label{EQ4}
\end{equation}
where $\boldsymbol{\mu}_n$ is the permanent dipole moment of state $\psi_n$ and $\boldsymbol{\alpha}(\omega)_n$ is the corresponding second-order frequency-dependent response, which we will refer to as polarizability for simplicity throughout. Note that Eq.~(\ref{EQ4}) corresponds to the dark-cavity limit, which yields a particularly clean functional form for analyzing the structure of $m$. The LFER construction, however, is not restricted to this limit. To illustrate its generality, we also consider a driven-field formulation based on the AC Stark effect in the Supporting Information Sec.~5. 

In this work, we neglect cavity-induced changes in entropic and thermal contributions, such that the changes in activation and reaction free energies are approximated by the corresponding electronic-energy shifts. We therefore define the change to activation energy as $\Delta\Delta G^\ddag \approx \Delta E_{\mathrm{TS}} - \Delta E_{\mathrm{R}}$ and the change to reaction energy as $\Delta\Delta G_{\mathrm{rxn}} \approx \Delta E_{\mathrm{P}} - \Delta E_{\mathrm{R}}$, the corresponding cavity-induced changes are given by

\begin{align}
\Delta\Delta G^{\ddag} &= \Delta E_{\mathrm{TS}} - \Delta E_{\mathrm{R}}   \label{change1}\\
&= \frac{1}{2}\boldsymbol{\lambda}^{2} \left( \boldsymbol{\mu}^2_{\mathrm{TS}} - \boldsymbol{\mu}^2_{\mathrm{R}} - \omega\left[ \boldsymbol{\alpha}(\omega)_{\mathrm{TS}} - \boldsymbol{\alpha}(\omega)_{\mathrm{R}} \right] \right),\nonumber \\
\Delta\Delta G_{\mathrm{rxn}} &= \Delta E_{\mathrm{P}} - \Delta E_{\mathrm{R}}  \label{change2}\\
&= \frac{1}{2}\boldsymbol{\lambda}^{2} \left( \boldsymbol{\mu}^2_{\mathrm{P}} - \boldsymbol{\mu}^2_{\mathrm{R}} - \omega\left[ \boldsymbol{\alpha}(\omega)_{\mathrm{P}} - \boldsymbol{\alpha}(\omega)_{\mathrm{R}} \right] \right). \nonumber 
\end{align}

Inserting Eqs.(\ref{change1},\ref{change2}) into Eq.\eqref{lin} yields the cavity-analogue of the LFER prefactor from Eq.~\eqref{prefactor} as

\begin{align}
m &= \frac{ \boldsymbol{\mu}^2_{\mathrm{TS}} - \boldsymbol{\mu}^2_{\mathrm{R}} -  \omega\left[ \boldsymbol{\alpha}(\omega)_{\mathrm{TS}} - \boldsymbol{\alpha}(\omega)_{\mathrm{R}} \right] }
{ \boldsymbol{\mu}^2_{\mathrm{P}} - \boldsymbol{\mu}^2_{\mathrm{R}} -  \omega\left[ \boldsymbol{\alpha}(\omega)_{\mathrm{P}} - \boldsymbol{\alpha}(\omega)_{\mathrm{R}} \right] }.
\label{eq:cavity-LFER}
\end{align}

It is important to note, and central to our approach, that although the perturbative energy shift depends on the light–matter coupling strength, the coupling strength itself cancels in $m$. The resulting prefactor therefore depends only on coupling-strength-independent molecular response properties and the cavity frequency, analogous to the field-free dipoles in the original electrostatic LFER. In the original LFER prefactor (Eq.~\eqref{prefactor}), the ratio of dipole moment differences can assume either positive or negative values, depending on the molecular dipoles involved. A similar situation arises in the cavity-modified prefactor (Eq.~\eqref{eq:cavity-LFER}), however, here the prefactor explicitly depends on the cavity frequency $\omega$ through the polarizability. This frequency dependence makes both the sign and magnitude of $m$ tunable, allowing the reaction to switch between BEP and Opposite-BEP behavior as the cavity mode is varied. Hence we can categorize cavity-modified reactions into four distinct types, Fig.~\ref{fig:figure0}(b).  
Type I: Reactions that, regardless of the cavity frequency, always follow the BEP relationship ($+m$). Type II: Reactions that switch from BEP ($+m$) to Opposite BEP ($-m$) behavior upon tuning the cavity frequency. Type III: Reactions that switch from Opposite BEP ($-m$) to BEP ($+m$) behavior. Type IV: reactions that, independent of the cavity tuning, consistently exhibit the counterintuitive Opposite BEP ($-m$) behavior.

Let us first examine the origin of these four reaction types through a functional analysis of the prefactor $m$. Since $m$ is constructed from differences of the cavity-induced energy shifts, we begin with the functional form of $\Delta E$. As shown in Fig.~\ref{fig:figure1}(a) and Eq.~(3), $\Delta E$ contains two contributions: a frequency-independent offset set by the permanent-dipole term and a frequency-dependent, pole-like contribution arising from the polarizability. The latter produces the characteristic $1/x$-type structure, with the pole position determined by the molecular resonance. The same structure carries over to the activation and reaction-terms that form the numerator and denominator of $m$, with their respective offset determined by $\Delta\boldsymbol{\mu}^{2\ddag}$ and $\Delta\boldsymbol{\mu}^2_{\mathrm{rxn}}$, and their pole-structure by $\omega\Delta\boldsymbol{\alpha}(\omega)^\ddag$ and $\omega\Delta\boldsymbol{\alpha}(\omega)_{\mathrm{rxn}}$. Note that the vacuum-induced second-order energy correction contains only the emission-like frequency branch (negative-frequency poles), whereas the conventional dynamic polarizability contains both positive- and negative-frequency branches. See Supporting Information Sec.~5 for further discussion.

\begin{figure*}[t]
    \centering
    \includegraphics[width=\linewidth]{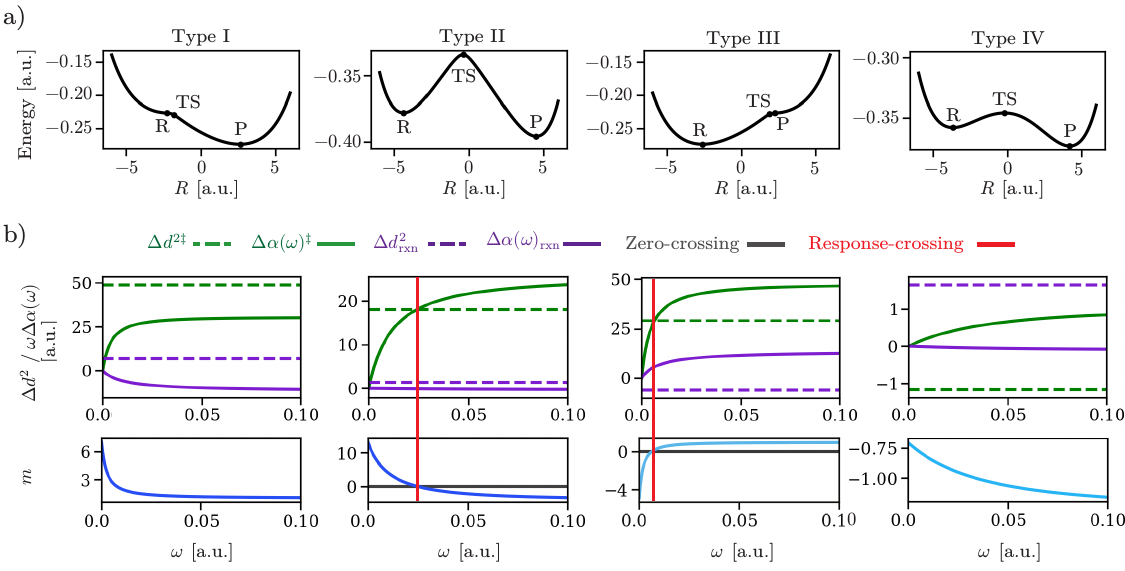}
\caption{\textbf{Exploring reaction types in the cavity-modified Shin--Metiu model.}
(a) Cavity-free ground-state Born--Oppenheimer energy surfaces for the four different parameter sets considered. R, TS, and P denote the reactant, transition state, and product, respectively.
(b) Top panels: numerator (green) and denominator (purple) contributions entering $m$ for the four reaction types, shown as a function of cavity frequency. The corresponding permanent-dipole contributions are shown as dashed lines and the polarizability contributions as solid lines. Bottom panels: resulting $m$ as a function of cavity frequency. The gray horizontal lines indicate zero-crossings of $m$, while the red vertical lines indicate the corresponding response crossings.}
    \label{fig:figure2}
\end{figure*}

This immediately provides a simple rule for the different reaction types, Fig.~\ref{fig:figure1}(b,c). If the polarizability contribution does not cross the corresponding permanent-dipole contribution over the range of cavity frequencies considered, the signs of the numerator and denominator remain unchanged. The relative signs of the numerator and denominator then determine whether $m$ remains positive (BEP) or negative (Opposite BEP) independent of cavity tuning (Fig.~\ref{fig:figure1}(b)). In contrast, when the polarizability contribution crosses the corresponding permanent-dipole contribution, the resulting numerator or denominator of $m$ changes sign at that cavity frequency (Fig.~\ref{fig:figure1}(c)). We will denote this dipole-polarizability-crossing $\Delta\boldsymbol{\mu}^2 = \omega \Delta\boldsymbol{\alpha}(\omega)$ as \textit{response-crossing} throughout the paper. Importantly, the resulting shape of $m$ depends on if the response crossing occurs in the numerator or denominator.  A numerator response-crossing, Fig.~\ref{fig:figure1}(c) first red line, yields a zero-crossing of $m$, as in the Type II and III behaviors shown in Fig.~\ref{fig:figure0}(b). A denominator response-crossing, on the other hand, corresponds to a vanishing denominator and therefore gives rise to a pole in $m$, Fig.~\ref{fig:figure1}(c) second red line. Interestingly, this pole does not arise from a molecular resonance, but solely from the response crossing in the denominator. However, similar to a resonance, $m$ becomes very large and highly sensitive to the cavity frequency near this pole. In this regime, the cavity can produce almost no net change in the reaction thermodynamics 
$\left(\Delta\boldsymbol{\mu}_{\mathrm{rxn}}^2 - \omega\Delta\boldsymbol{\alpha}(\omega)_{\mathrm{rxn}} \approx 0\right)$ 
while still yielding a finite change in the reaction kinetics 
$\left(\Delta\boldsymbol{\mu}^{2\ddag} - \omega\Delta\boldsymbol{\alpha}(\omega)^\ddag \neq 0\right)$. Finally, if numerator and denominator response-crossings occur at different cavity frequencies, $m$ can exhibit a smooth zero-crossing (Type II/III) at one frequency and a pole at the other. Thus, the four reaction types can be understood directly in terms of response-crossings arising from the competition between the permanent-dipole and polarizability contributions.

\begin{figure*}[h!]
    \centering
    \includegraphics[width=\linewidth]{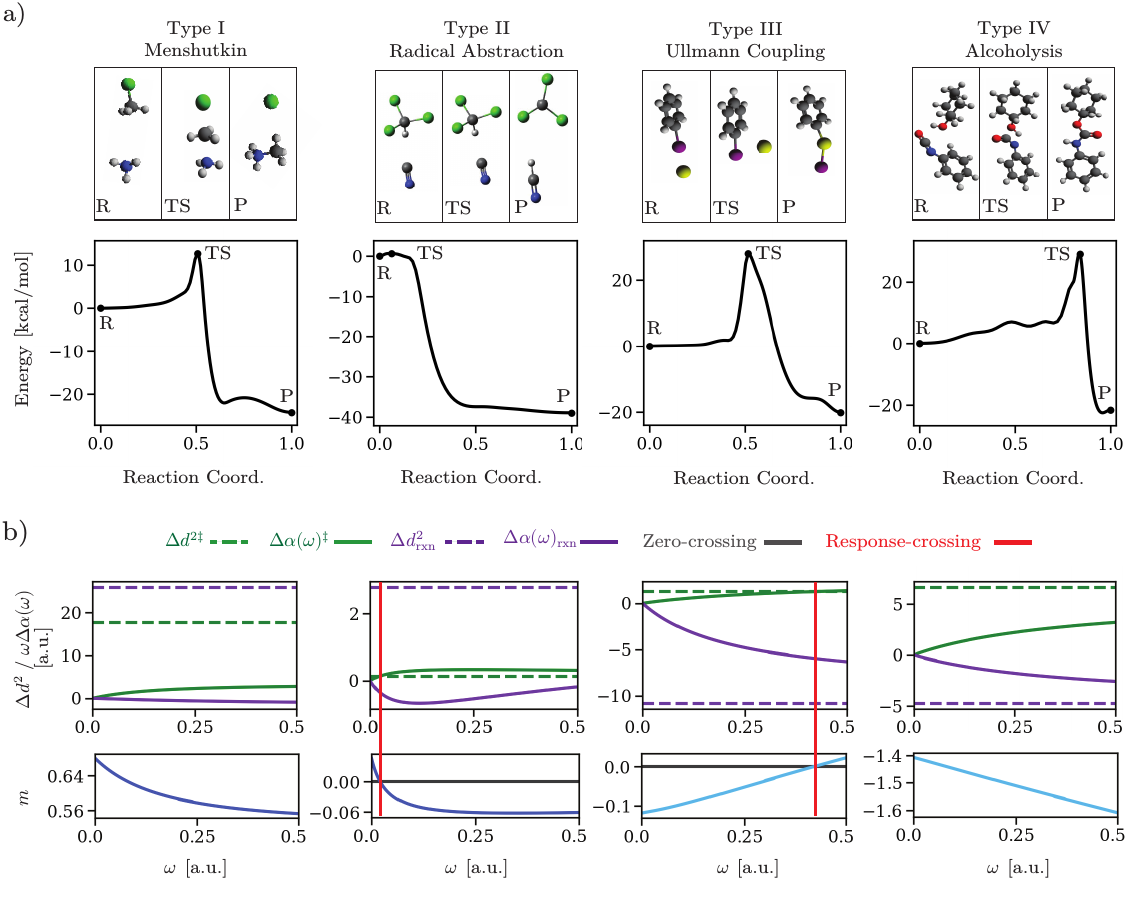}
\caption{\textbf{Exploring reaction types in cavity-modified reactions from ab initio calculations.}
(a) Cavity-free reaction pathways for the four reactions considered: the Type I Menshutkin reaction, Type II radical abstraction reaction, Type III dehalogenation step of the Ullmann coupling reaction, and Type IV alcoholysis reaction. R, TS, and P denote the reactant, transition state, and product, respectively.
(b) Top panels: numerator (green) and denominator (purple) contributions entering $m$ for each reaction, shown as a function of cavity frequency. The corresponding permanent-dipole contributions are shown as dashed lines and the polarizability contributions as solid lines. Bottom panels: resulting $m$ as a function of cavity frequency. The gray horizontal lines indicate a zero-crossing of $m$, while the red vertical lines indicate the corresponding response crossings.}
    \label{fig:figure3}
\end{figure*}

To close this analytical discussion, we briefly consider the generality of the framework. As discussed above, the common light–matter coupling-strength prefactor cancels from $m$, making the reaction-type classification independent of coupling strength. However, the present formulation is perturbative, consistent with the LFER assumption that the reaction pathway is not substantially altered by the cavity, particularly in collective regimes where the coupling per molecule is comparatively weak \autocite{herrera2016, feist2017}. Perturbative treatments can also remain accurate beyond the weak-coupling limit in suitable representations, such as a coherent-state basis \autocite{roden2024coherent}. For all systems considered here, we benchmark the adjusted expression against full second-order perturbation theory and exact solutions, restricting our analysis to the regime where it remains valid (Supporting Information Figs.~S1--S3). Additionally, the framework does not rely on a particular number of molecules or degree of collectivity, but only on the resulting, coupling-strength-independent dipole and polarizability contributions that enter $m$. These quantities can be obtained from a single molecule \autocite{pavosevic2023nature}, many-molecule calculations \autocite{li2020cavity, lindoy2024dynamics}, \textit{ab initio} methods \autocite{flick2015, flick2018oep, pellegrini2015}, or effective descriptions that incorporate local-field or hotspot effects \autocite{wickramasinghe2026mode, sidler2026}. Hence, $m$ provides a general and coarse-grained way to anticipate how cavity coupling can affect reactivity from these response properties alone.

To test this analytical classification, we next consider two complementary settings: the Shin--Metiu model (Fig.~\ref{fig:figure2}) as a controlled benchmark and a set of molecular reactions calculated with \textit{ab-initio} electronic-structure methods (Fig.~\ref{fig:figure3}). We begin with the Shin--Metiu model\autocite{shin1995}, a one-dimensional proton-coupled electron-transfer model with one moving proton and electron between two fixed ions. The interactions are described by soft-Coulomb potentials, whose parameters are varied to explore different reaction profiles and response properties. Additional details of the Hamiltonian \autocite{shin1995, Fang1997, Fang1997PCET, Hoffmann2020ManyModes}, model parameters, and light-matter coupling-strength benchmarks are provided in the Supporting Information Sec.~3. Within the Born--Oppenheimer approximation, we calculate the electronic states, permanent dipoles, transition dipoles, and corresponding frequency-dependent polarizabilities, which are then used to calculate the cavity-modified prefactor $m$ (Eq.(\ref{eq:cavity-LFER})). Fig.~\ref{fig:figure2}(a) shows the reaction pathway for four different parameter sets. Fig.~\ref{fig:figure2}(b) shows the corresponding activation (green) and reaction (purple) dipole contributions (dashed) and polarizability contributions (solid), while the bottom panel of Fig.~\ref{fig:figure2}(b) shows the resulting prefactor $m$ (blue).

Using the response-crossing rules introduced above, we can directly predict the four distinct reaction types from the model response properties. More precisely, for column one and four in Fig.~\ref{fig:figure2}, no response crossing occurs over the cavity-frequency range considered, and the signs of the numerator and denominator therefore remain fixed. Hence we can classify these two reactions as Types I and IV. In Type I, the corresponding numerator and denominator have the same sign, yielding $m>0$ and a BEP behavior for all cavity frequencies. In Type IV, the corresponding numerator and denominator have opposite signs, yielding $m<0$ and a Opposite BEP behavior for all cavity frequencies. In contrast, column two and three in Fig.~\ref{fig:figure2} do exhibit a response-crossing in the activation contribution, i.e., in the numerator of $m$, which produces a zero-crossing in $m$. Hence we can classify these two reactions as Types II and III. For Type II $m$ is positive at low frequency and the reaction switches from BEP to Opposite BEP, whereas for Type III $m$ is initially negative and switches from Opposite BEP to BEP. Thus, the four reaction types emerge directly from whether a response-crossing occurs and from the relative signs of the dipole and polarizability contributions, providing a direct numerical realization of the analytical picture developed above.

Having established the four reaction types in the Shin--Metiu model, we next turn to molecular reactions calculated \textit{ab-initio} with electronic-structure methods. We consider four representative reactions: the Menshutkin reaction \autocite{climent2019plasmonic}, a radical abstraction reaction \autocite{fidler2023}, the dehalogenation step of a Ullmann coupling reaction \autocite{hoffmann2022linear}, and an alcoholysis reaction \autocite{ahn2023science}. All calculations are performed in ORCA 4.2.1 \autocite{neeseORCA, neeseORCA2018} using B3LYP/6-31G(d,p) with water, B3LYP/Def2-TZVP, CAM-B3LYP/Def2-TZVP, and CAM-B3LYP/6-31G(d,p) with tetrahydrofuran, respectively. For each reaction, the reactant and product structures are optimized and the transition state is obtained using nudged elastic band calculations. The resulting electronic states, permanent dipoles and transition dipoles are then used to calculate the cavity-modified prefactor $m$ (Eq.(\ref{eq:cavity-LFER})). Additional computational details, workflows, and benchmarks are provided in Supporting Information Sec.~4.

Fig.~\ref{fig:figure3}(a) shows the four reaction pathways, Fig.~\ref{fig:figure3}(b) shows the corresponding activation (green) and reaction (purple) dipole contributions (dashed) and polarizability contributions (solid), and Fig.~\ref{fig:figure3}(b) bottom panel shows the resulting prefactor $m$. As in the Shin--Metiu model, each molecular reaction can be simply classified as one reaction type through their response-crossings. The Menshutkin and alcoholysis reactions exhibit no response-crossing over the cavity-frequency range considered and therefore no zero-crossing in $m$. Hence they can be classified as Type I and IV reactions. For the Menshutkin reaction, the numerator and denominator retain the same sign across the cavity-frequency range, yielding $m>0$, corresponding to BEP behavior and therefore a Type I reaction. For the alcoholysis reaction, the numerator and denominator retain the opposite sign across the cavity-frequency range, yielding $m<0$, corresponding to Opposite-BEP behavior, and therefore a Type IV reaction. By contrast, the radical abstraction and dehalogenation process both exhibit a response-crossing in the numerator of $m$, yielding a zero-crossing in $m$. Hence they can be classified as Type II and III reactions. For the radical abstraction, $m$ is positive at low frequency and hence changes from BEP to Opposite BEP at the response-crossing, corresponding to a Type II reaction. While for the dehalogenation process, $m$ is initially negative and switches from Opposite BEP to BEP, corresponding to Type III reaction. Thus, the same response-crossing picture that emerges analytically and in the Shin--Metiu model also classifies the behavior of chemically realistic reactions. Importantly, this classification applies across very different reaction classes and model systems, showing that cavity-modified reactivity can exhibit qualitatively distinct and even counterintuitive behavior that can nevertheless be understood from simple molecular response properties.\\

\section*{Conclusion and Outlook}

In conclusion, polaritonic chemistry remains a complicated field, with many different answers to the same question of how a cavity modifies chemical reactivity. Here, we deliberately take a more zoomed-out view of polaritonic chemistry inspired by electrostatic catalysis. To this end, we extend the nonempirical linear free-energy relationship previously explored for reactions in static electric fields to cavity-modified chemistry. This allows us to establish simple response-crossing rules for how reactions can behave inside a cavity and to classify them into four distinct reaction types.

More precisely, Type I and Type IV reactions exhibit no response-crossing and hence follow either BEP or Opposite-BEP behavior regardless of the cavity frequency. Type II and Type III reactions instead exhibit a response-crossing and hence can switch from BEP to Opposite BEP, or vice versa, as the cavity frequency is tuned. Interestingly, if the response-crossing occurs in the denominator, $m$ can exhibit a pole at this cavity frequency, making the reaction highly sensitive to cavity-frequency tuning near the crossing. We demonstrate these four reaction types first in the Shin--Metiu model and then for four different reactions treated with \textit{ab initio} electronic-structure methods. In all cases, the reaction behavior follows the same simple response-crossing rules, governed by the competition between permanent-dipole and polarizability contributions.

This framework provides a deliberately coarse-grained way to think about polaritonic chemistry by focusing on the response properties that enter $m$. These quantities can be obtained from single molecules, many-molecule calculations, model systems, \textit{ab initio} methods, or effective descriptions that include local-field or hotspot effects. Interestingly, even at this coarse-grained level, cavity-modified reactivity can already exhibit qualitatively different and strongly counterintuitive behavior. A reaction can follow conventional BEP behavior at one cavity frequency and Opposite-BEP behavior at another. Yet this seemingly complicated behavior can still be understood from a simple response-crossing picture. In this way, the reaction-type classification provides a general way to organize and anticipate the different forms of reactivity that can emerge in cavities.

\section*{Acknowledgements}

Support was provided in part through NYU IT High Performance Computing resources, services, and staff expertise. Simulations were partially executed on resources supported by the SCCPC at NYU.

\section*{Funding}

This work was supported by the Simons Center for Computational Physical Chemistry (SCCPC) at NYU (SF Grant No. 839534) and the U.S. DOE, Office of Science, BES (Early Career Award No. DE-SC0026328). J.G.H. acknowledges support from a graduate fellowship, and D.M.W. acknowledges support from a postdoctoral fellowship, both from the Simons Foundation [MPS-T-MPS-00839534, MET].

\section*{Competing Interests}

The authors declare no competing interests.

\section*{Supporting information}

Supporting Information includes theoretical derivations, computational details and convergence analyses for the perturbative treatment, Shin--Metiu model, and \textit{ab initio} reactions, as well as the extension of the LFER framework to AC Stark shifts.

\section*{Data Availability}
Code and data for reproducing the calculations and figures in this manuscript can be found at this \href{https://github.com/HoffmannGroup-NYU/Exploring-Reaction-Types}{Github Repository}.

\printbibliography

\newpage

\begin{figure}[h!]
    \centering
    \includegraphics[width=3.25in]{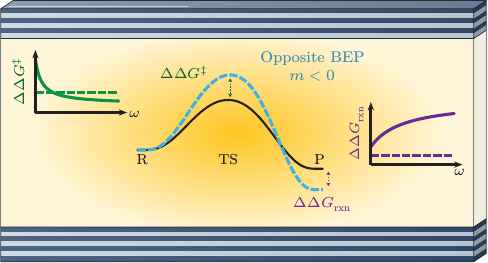}
    \caption*{For Table of Contents Only}
    \label{fig:TOC}
\end{figure}

\end{document}